\documentclass[aps,pra,twocolumn,showpacs,superscriptaddress,groupedaddress]{revtex4}  
\usepackage{graphicx}  
\usepackage{dcolumn}   
\usepackage{bm}        
\usepackage{amssymb,amsmath,amsopn,amsfonts}
\usepackage{colordvi}
\usepackage{algorithm}
\usepackage{algpseudocode}
\usepackage[sort&compress,square]{}
\usepackage{color}

\begin{document}

\widetext



\title{Sampling-based learning control of inhomogeneous quantum
ensembles}
\author{Chunlin~Chen}
\affiliation{Department of Control and System Engineering, Nanjing
University, Nanjing 210093, China}
\affiliation{Department of
Chemistry, Princeton University, Princeton, New Jersey 08544, USA}
\author{Daoyi~Dong}\email{daoyidong@gmail.com}
\affiliation{School of Engineering and Information Technology, University of New South Wales
at the Australian Defence Force Academy,
Canberra, ACT 2600, Australia}
\author{Ruixing~Long}
\affiliation{Department of
Chemistry, Princeton University, Princeton, New Jersey 08544, USA}
\author{Ian R.~Petersen}
\affiliation{School of Engineering and Information Technology, University of New South Wales
at the Australian Defence Force Academy,
Canberra, ACT 2600, Australia}
\author{Herschel A.~Rabitz}
\affiliation{Department of
Chemistry, Princeton University, Princeton, New Jersey 08544, USA}

\date{\today}

\begin{abstract}
Compensation for parameter dispersion is a
significant challenge for control of inhomogeneous quantum ensembles. In this
paper, we present a systematic methodology of sampling-based
learning control (SLC) for simultaneously steering the members of
inhomogeneous quantum ensembles to the same desired state. The SLC
method is employed for optimal control of the
state-to-state transition probability for inhomogeneous quantum ensembles
of spins as well as $\Lambda$ type atomic systems.
The procedure involves the steps of (i) training and (ii) testing. In the training step, a generalized system is
constructed by sampling members according to the
distribution of inhomogeneous parameters drawn from the ensemble. A gradient flow based
learning and optimization algorithm is adopted to find the control
for the generalized system. In the process of testing, a number of additional ensemble members are randomly selected to
evaluate the control performance. Numerical results
are presented showing the success of the SLC
method.
\end{abstract}

\maketitle


\section{Introduction}\label{Sec1}
Control of quantum phenomena lies at the heart of emerging quantum technology
\cite{Dong and Petersen 2010IET}-\cite{Brif et al 2010}. Quantum
control theory has many components including controllability assessment, optimal control, feedback control,
etc. Most existing results focus on control
design of single quantum systems \cite{Dong and Petersen
2010IET}-\cite{Judson and Rabitz 1992}. Another important issue is control design for quantum
ensembles. A quantum ensemble consists of a large number of (up to
$\sim 10^{23}$) single quantum systems (e.g., spin systems) and every quantum system is referred to as
a member of the ensemble in this paper.
Quantum ensembles have wide applications in emerging quantum
technology including quantum computation \cite{Cory et al 1997},
long-distance quantum communication \cite{Duan et 2001}, quantum memory \cite{Bensky et al 2012}, and
magnetic resonance imaging \cite{Li et al 2011PNAS}. Several results
on quantum ensemble control have been presented including unitary control in
homogeneous quantum ensembles for maximizing signal
intensity in coherent spectroscopy \cite{Glaser et al 1998} and
feedback stabilization of quantum ensembles \cite{Altafini 2007-1}.

In practical applications, the members of a quantum ensemble could
have variations in the parameters that characterize the system
dynamics \cite{Levitt 1986}, \cite{Li and Khaneja 2006}. For
example, the spins of an ensemble in nuclear magnetic resonance (NMR)
experiments may encounter large dispersion in the strength of the applied
radio frequency field (rf imhomogeneity) as well as the members exhibiting variations in their natural
frequencies (Larmor dispersion) \cite{Skinner et al 2003}, \cite{Khaneja et al 2005}. In this paper, these situations are referred as {\it inhomogeneous} quantum ensembles. It is
generally impractical to employ different control inputs for
individual members of a quantum ensemble in the laboratory.
Hence, it is important to develop the means for designing
control fields that can simultaneously steer the ensemble
of systems from an initial state to a desired target state when
variations exist in the system parameters. Such controls
are also called compensating pulse sequences in NMR spectroscopy
\cite{Levitt 1986}, \cite{Vandersypen and Chuang 2004}. Other applications include control of a randomly oriented ensemble of molecules in physical chemistry \cite{Turinici and Rabitz 2004}, the design of slice
selective excitation and inversion pulses in magnetic resonance
imaging, and the correction of systematic errors in quantum information
processing \cite{Li and Khaneja 2006}. Theoretical results showed that under commonly arising conditions there exist
optimal laser fields to control all molecules in an inhomogeneous ensemble, regardless of their orientation or spatial location \cite{Turinici et al 2004JPA}, \cite{Rabitz and Turinici 2007}. Recent studies considered the
controllability and optimal control of inhomogeneous spin ensembles \cite{Li et al 2011PNAS},  \cite{Li and
Khaneja 2006}, \cite{Li and Khaneja 2009}-\cite{Owrutsky and Khaneja 2012}. An additional investigation considered the stabilization
of an inhomogeneous ensemble of non-interacting spin systems
using Lyapunov control methodology \cite{Beauchard et al 2012}.

This paper presents a systematic methodology for control
design of inhomogeneous quantum ensembles for the
state-to-state transition probability, specifically for spins
and three-level $\Lambda$ type systems.
The proposed method involves the steps of (i) training and (ii) testing, and we call sampling-based learning control
(SLC). In the training step, we sample several members according to
the distribution of inhomogeneous parameters from the ensemble and construct a
generalized system using these collective samples. Then we employ a gradient
flow based learning and optimization algorithm \cite{Long and Rabitz 2011} to find the control
providing good performance for the generalized system. In the
process of testing the deduced controls, we randomly select a number of
sampling members to evaluate the control performance. Numerical
simulations show that the SLC method has potential for
 practical control design of various inhomogeneous quantum
ensembles. These findings support the previous theoretical analysis suggesting that control on inhomogeneous ensembles should generally be feasible \cite{Turinici and Rabitz 2004}-\cite{Li and Khaneja 2009}.

The paper is organized as follows. Section \ref{Sec2} formulates
the control problem for inhomogeneous quantum ensembles and presents the details of SLC. The SLC method is illustrated for a
two-level inhomogeneous quantum ensemble in Section \ref{Sec3}, and for a three-level
inhomogeneous quantum ensemble in Section \ref{Sec4}. Conclusions are
presented in Section \ref{Sec5}.

\section{Methodology}\label{Sec2}

\subsection{Model and problem formulation}
Consider a finite-dimensional closed quantum system where the
evolution of its state $|\psi(t)\rangle$ is described by the
Schr\"{o}dinger equation (setting $\hbar=1$):
\begin{equation} \label{systemmodel}
\left\{ \begin{array}{l}
  \frac{d}{dt}|{\psi}(t)\rangle=-iH(t)|\psi(t)\rangle \\
 t\in [0, T], \ |\psi(0)\rangle=|\psi_{0}\rangle.\\
\end{array}
\right.
\end{equation}
The solution of (\ref{systemmodel}) is given
by $\displaystyle |\psi(t)\rangle=U(t)|\psi_{0}\rangle$, where the
propagator $U(t)$ satisfies
\begin{equation}
\left\{ \begin{array}{c}
  \frac{d}{dt}U(t)=-iH(t)U(t),\\
  t\in [0, T], \ U(0)=\textrm{Id}.\\
\end{array}
\right.
\end{equation}

In this paper, we consider an inhomogeneous ensemble in which the Hamiltonian of
each member has the following form
\begin{equation}\label{controlmodel}
H_{\omega, \theta}(t)=g(\omega)H_{0}+b(\theta)\sum_{m=1}^{M}u_{m}(t)H_{m},
\end{equation}
where $H_{0}$ is the free Hamiltonian and
$\sum_{m=1}^{M}u_{m}(t)H_{m}$ corresponds to the time-dependent control
Hamiltonian that represents the interaction of the system with the
external fields $u_{m}(t)$ (real-valued and square-integrable
functions) through Hermitian operators $H_{m}$. The functions $g(\omega)$ and $b(\theta)$ characterize
the inhomogeneous distribution in the free Hamiltonian and
control Hamiltonian, respectively (see Fig. 1). In this paper, we assume that $g(\omega)=\omega$ and $b(\theta)=\theta$, and the parameters
$\omega$ and $\theta$  are time independent and uniformly distributed over $[1-\Omega, 1+\Omega]$ and $[1-\Theta, 1+\Theta]$, respectively. The constants
$\Omega \in [0,1]$ and $\Theta \in [0,1]$ represent the bounds of
the parameter dispersion. The objective is to design the controls
$\{u_{m}(t), m=1,2,\ldots , M\}$ to simultaneously drive the
members (with different $\omega$ and $\theta$) of the quantum
ensemble from an initial state $|\psi_{0}\rangle$ to the same target
state $|\psi_{\text{target}}\rangle$ with high fidelity. The control
outcome is described by a \emph{performance function} $J(u)$ for
each control strategy $u=\{u_{m}(t), m=1,2,\ldots , M\}$. The
control problem can then be formulated as a maximization problem as
follows:
\begin{equation}\label{ensemble control}
\begin{split}
\displaystyle \max_u  & J(u):=\max_u \mathbb{E}[J_{\omega,\theta}(u)]\\ 
\text{s.t.} \ \ \ & \frac{d}{dt}|\psi_{\omega,\theta}(t)\rangle=-iH_{\omega,\theta}(t)|\psi(t)\rangle,~ t \in [0, T], \\
& |\psi_{\omega,\theta}(0)\rangle=|\psi_{0}\rangle \\
& H_{\omega,\theta}(t)=\omega H_{0}+\theta \sum_{m=1}^{M}u_{m}(t)H_{m}, \\
& \omega \in [1-\Omega,1+\Omega], \ \ \theta \in [1-\Theta,1+\Theta],
\end{split}
\end{equation}
where $J_{\omega,\theta}(u)$ is a fidelity measure of each
member of the ensemble and $\mathbb{E}[J_{\omega,\theta}(u)]$
denotes the average value of $J_{\omega,\theta}$ over the
ensemble. The fidelity between the final state
$|\psi_{\omega,\theta}(T)\rangle$ and the target state
$|\psi_{\text{target}}\rangle$ is defined as follows \cite{Nielsen
and Chuang 2000}
\begin{equation}\label{fidelity}
F(|\psi_{\omega,\theta}(T)\rangle,|\psi_{\text{target}}\rangle)=|\langle \psi_{\omega,\theta}(T)|\psi_{\text{target}}\rangle| .
\end{equation}
The fidelity $F$ is used to evaluate the performance of a designed control in the testing step. However, for convenient calculation of a gradient flow
in the training step, we take the performance function $J(u)=F^{2}$; i.e.,
$$J_{\omega,\theta}(u):=\vert
\langle\psi_{\omega,\theta}(T)|\psi_{\text{target}}\rangle\vert^{2}.$$
Note that $J_{\omega,\theta}$ depends implicitly on the control
$u$ through the Schr\"odinger equation.

\subsection{Sampling-based learning control of inhomogeneous quantum ensembles}
Gradient-based methods \cite{Brif et al 2010}, \cite{Long and Rabitz
2011}, \cite{Jirari and Potz 2005}, \cite{Roslund and Rabitz 2009}
have been successfully applied to search for optimal solutions to a
variety of quantum control problems, including the theoretical and
laboratory applications. In this paper, a gradient-based learning
method is employed to optimize the controls for inhomogeneous
quantum ensembles.
We present a systematic methodology for ensemble control design
utilizing selected samples (as shown in Fig. 1) from the ensemble.
These samples are drawn from the distribution of inhomogeneous parameters to design the control. Then the
resultant control is applied to additional ensemble members to test the control performance. As such, the SLC method includes the steps of
(i) training and (ii) testing.

\begin{figure}
\centering
\includegraphics[width=3.5in]{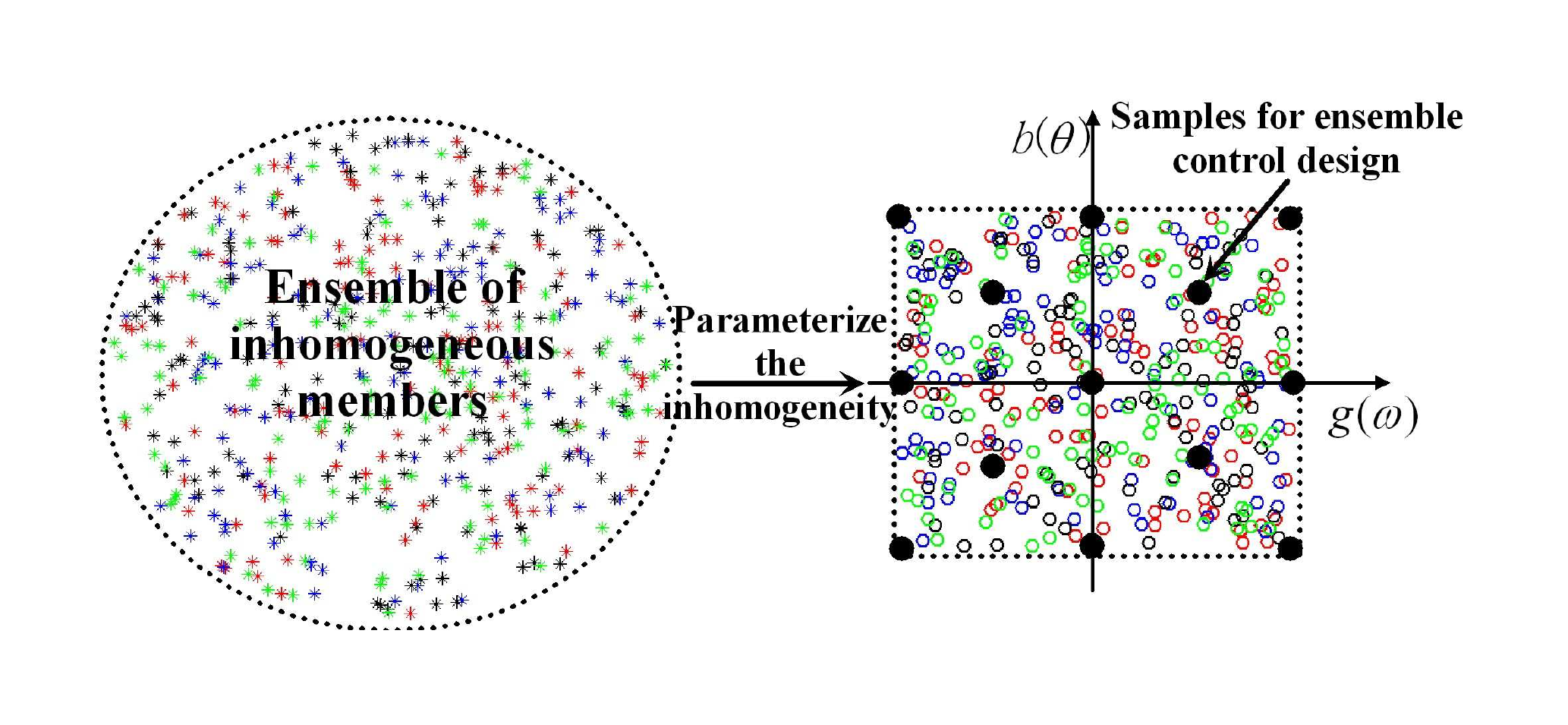}
\caption{(color online) A schematic for modeling an inhomogeneous ensemble
parameterized with $g(\omega)$ and $b(\theta)$, and the selection of
samples to construct a generalized system. An ensemble of inhomogeneous members
on the left is mapped into a space determined by $g(\omega)$ and $b(\theta)$ on the right, where $g(\omega)$ and $b(\theta)$ characterize
the distribution of inhomogeneity in the free Hamiltonian and
control Hamiltonian, respectively. Samples for ensemble control design are
drawn from the distribution of inhomogeneous parameters. These samples are used to construct a generalized
system for learning a control field with the performance as an average on the samples.}
\end{figure}

\subsubsection{Training}\label{sec:training}
In the training step, we select $N$ sampled members from the quantum
ensemble according to the distribution (e.g., uniform
distribution) of the inhomogeneous parameters and then construct a
generalized system as follows
\begin{equation}\label{generalized-system}
\frac{d}{dt}\left(%
\begin{array}{c}
  |{\psi}_{\omega_1,\theta_1}(t)\rangle \\
  |{\psi}_{\omega_2,\theta_2}(t)\rangle \\
  \vdots \\
  |{\psi}_{\omega_N,\theta_N}(t)\rangle \\
\end{array}%
\right)
=-i\left(%
\begin{array}{c}
  H_{\omega_1,\theta_1}(t)|\psi_{\omega_1,\theta_1}(t)\rangle \\
  H_{\omega_2,\theta_2}(t)|\psi_{\omega_2,\theta_2}(t)\rangle \\
  \vdots \\
  H_{\omega_N,\theta_N}(t)|\psi_{\omega_N,\theta_N}(t)\rangle \\
\end{array}%
\right),
\end{equation}
where
$H_{\omega_n,\theta_n}=\omega_{n} H_{0}+\theta_{n} \sum_{m}u_{m}(t)H_{m}$
with $n=1,2,\dots,N$. The performance function for the generalized
system is defined by
\begin{equation}\label{eq:cost}
J_N(u):=\frac{1}{N}\sum_{n=1}^N J_{\omega_n,\theta_n}(u)=\frac{1}{N}\sum_{n=1}^{N}\vert \langle\psi_{\omega_n,\theta_n}(T)|\psi_{\text{target}}\rangle\vert^{2}.
\end{equation} The goal of the training step is to find a control $u^*$
that maximizes the performance function defined in Eq. \eqref{eq:cost}.
The performance function is $J_N(u^{0})$ with an initial
control $u^{0}=\{u^{0}_{m}(t)\}$. We apply the
gradient flow method \cite{Brif et al 2010}, \cite{Khaneja et al 2005}, \cite{Long and Rabitz 2011}-\cite{Roslund and Rabitz 2009} to seek an optimal control
$u^{*}=\{u^{*}_{m}(t)\}$. The detailed gradient flow algorithm is provided in the Appendix (\emph{Algorithm 1}).
The time interval $[0,T]$ is divided equally into $Q$ time
slices $\triangle t$ and we assume that the controls are constant
within each time slice. The time
index is $t_{q}=qT/Q$, where $Q=T/\triangle t$ and
$q=1,2,\ldots,Q$.

The
motivation behind SLC is to design the
control using a minimal number of sample members. Therefore, it is
necessary to choose a representative set of samples.
For example, when the distributions of both $\omega$ and $\theta$
are uniform, we may choose some equally spaced samples in the
$\omega-\theta$ space. In this case, the intervals of $[1-\Omega,
1+\Omega]$ and $[1-\Theta, 1+\Theta]$ are divided into
$N_{\Omega}+1$ and $N_{\Theta}+1$ subintervals, respectively,
where $N_{\Omega}$ and $N_{\Theta}$ are conveniently chosen positive odd
integers. Then the total number of samples is $N=N_{\Omega}N_{\Theta}$,
where $\omega_{n}$ and $\theta_{n}$ are chosen from all
combinations of $(\omega_{n_{\Omega}}, \theta_{n_{\Theta}})$ as
follows
\begin{equation}\label{discrete}
\left\{ \begin{array}{c} \omega_{n} \in
\{\omega_{n_{\Omega}}=1-\Omega+\frac{(2n_{\Omega}-1)\Omega}{N_{\Omega}},
\ n_{\Omega}=1,2,\ldots, N_{\Omega}\},\\
\theta_{n} \in
\{\theta_{n_{\Theta}}=1-\Theta+\frac{(2n_{\Theta}-1)\Theta}{N_{\Theta}},\
\
n_{\Theta}=1,2,\ldots, N_{\Theta}\}. \\
\end{array}
\right.
\end{equation}

\subsubsection{Testing}
For testing, we apply the optimal
control $u^{*}$ obtained in the training step to additional
samples randomly selected  from the inhomogeneous quantum ensemble
and evaluate the control performance of each sample in terms of
the fidelity $F(|\psi(T)\rangle,|\psi_{\text{target}}\rangle)$ between the final state achieved by each sample
$|\psi(T)\rangle$ and the target state
$|\psi_{\text{target}}\rangle$.
If both the average value and the minimum value of the fidelity
$F(|\psi(T)\rangle,|\psi_{\text{target}}\rangle)$ for all the
tested samples are satisfactory, we accept the designed control
law and end the control design process. Otherwise, we return to the training step and generate another optimized control
strategy (e.g., restarting the training step with a new initial
control strategy or a new set of samples guided by the performance of the tested members).

\section{Control of two-level inhomogeneous quantum ensembles}\label{Sec3}
In this section, we apply SLC to two-level inhomogeneous ensembles. Several groups of numerical
experiments are given to evaluate the performance of SLC.

\subsection{Two-level inhomogeneous ensemble}
Consider a quantum ensemble consisting of two-level
quantum systems (e.g., spins). The Pauli
matrices $\sigma=(\sigma_{x},\sigma_{y},\sigma_{z})$ are denoted as follows:
\begin{equation}
\sigma_{x}=\begin{pmatrix}
  0 & 1  \\
  1 & 0  \\
\end{pmatrix} , \ \ \ \
\sigma_{y}=\begin{pmatrix}
  0 & -i  \\
  i & 0  \\
\end{pmatrix} , \ \ \ \
\sigma_{z}=\begin{pmatrix}
  1 & 0  \\
  0 & -1  \\
\end{pmatrix} .
\end{equation}
We let the free Hamiltonian be
$H_{0}=\frac{1}{2}\sigma_{z}$ and its two eigenstates are denoted as
$|0\rangle$ and $|1\rangle$. The control Hamiltonian is
$H_{u}=\frac{1}{2}u_{1}(t)\sigma_{x}+\frac{1}{2}u_{2}(t)\sigma_{y}$.
Then we have
\begin{equation}\label{generalmodel}
|\dot{\psi}(t)\rangle=-iH(t)|\psi(t)\rangle,
\end{equation}
where
$H(t)=H_{0}+H_{u}(t)=\frac{1}{2}\sigma_{z}+\frac{1}{2}u_{1}(t)\sigma_{x}+\frac{1}{2}u_{2}(t)\sigma_{y}$.
For the inhomogeneous ensemble, the Hamiltonian of each member is described as
\begin{equation}
H_{\omega,\theta}(t)=\omega H_{0}+\theta H_{u}(t).
\end{equation}
The state of the quantum system can be represented as
$|\psi(t)\rangle=c_{0}(t)|0\rangle+c_{1}(t)|1\rangle$. Denote
$C(t)=(c_{0}(t),c_{1}(t))^{T}$, where
$c_0(t)$ and $c_1(t)$ are complex amplitudes. We have
\begin{equation}
i\dot{C}(t)=(H_{0}+H_{u}(t))C(t).
\end{equation}

To construct a generalized system for the training step, we select $N$ members
($n=1, 2, \ldots, N$) from the ensemble as follows:
\begin{equation}\label{element2level}
\left(%
\begin{array}{c}
   \dot{c}_{0,n}(t) \\
   \dot{c}_{1,n}(t) \\
\end{array}%
\right)=
\left(%
\begin{array}{cc}
  0.5\omega_{n} i & \theta_{n} f(u) \\
  -\theta_{n} f^{*}(u) & -0.5\omega_{n} i \\
\end{array}%
\right) \left(%
\begin{array}{c}
   c_{0, n}(t)  \\
  c_{1, n}(t) \\
\end{array}%
\right),
\end{equation}
where $f(u)=u_{2}(t)-0.5iu_{1}(t)$, $\omega_{n} \in [1-\Omega, 1+\Omega]$ and $\theta_{n} \in
[1-\Theta, 1+\Theta]$ have uniform distributions. The objective
is to find a control $u(t)=\{u_{m}(t), m=1,2\}$ to drive
all the inhomogeneous members from an initial state
$|\psi_{0}\rangle=|0\rangle$; i.e., $C_{0}=(1,0)^T$, to the target
state $|\psi_{\text{target}}\rangle=|1\rangle$; i.e.,
$C_{\textrm{target}}=(0,1)^T$. We construct a generalized system
for the training samples using Eq. (\ref{generalized-system})
with the performance function $J_{N}(u)$ in Eq. (\ref{eq:cost}).

The task is to find the control $u(t)$ to maximize the
performance function $J_{N}(u)$. For a given small
threshold $\epsilon>0$, if $J_{N}(u)>1-\epsilon$, we find
a suitable candidate control law for the generalized system.
We employ \emph{Algorithm 1} to find the optimal control
$u^{*}(t)=\{u^{*}_{m}(t), m=1,2\}$ for this generalized system.
This optimal control is then applied to other randomly
selected members to test its performance.

\subsection{Numerical results}
Several groups of numerical
experiments are carried out on an inhomogeneous spin
ensemble to demonstrate SLC. The parameter settings are as follows: $\Omega=0.2$ and $\Theta=0.2$; the target time is $T=2$
and the total time interval $[0,T]$ is divided equally into
$Q=200$ time steps, $\Delta
t=T/Q=0.01$; the learning rate is set as
$\eta^{k}=0.2$; the control strategy is initialized as
$u^{k=0}(t)=\{u^{0}_{1}(t)=\sin t, u^{0}_{2}(t)=\sin t\}$.

\subsubsection{Performance with two controls}
To construct a generalized system for the inhomogeneous ensemble with
parameter dispersion on both $\omega$ and $\theta$, we choose
$N_{\Omega}=5$ and $N_{\Theta}=5$ such that
$N=N_{\Omega}N_{\Theta}=25$ samples are employed in the learning phase. Using Eq.
(\ref{discrete}), we have
\begin{equation}\label{discrete2}
\left\{ \begin{split}
& \omega_{n}=1-0.2+\frac{0.2(2\text{fix}(n/5)-1)}{5},\\
& \theta_{n}=1-0.2+\frac{0.2(2(\textrm{mod}(n,5)-1)}{5}, \\
\end{split}\right.
\end{equation}
where $n=1,2,\ldots,25$, $\text{fix}(x)=\max \{z\in \mathbb{Z}|z\leq x\}$, $\text{mod}(n,5)=n-5z\ (z\in \mathbb{Z}\ \text{and}\ \frac{n}{5}-1<z\leq \frac{n}{5} )$ and $\mathbb{Z}$ is the set of integers. We set $\epsilon=5\times 10^{-5}$. The
learned optimal control strategy is given as in Fig. 2 and the
testing performance in Fig. 3 shows that the fidelities for the state
transition lie in the interval of $[0.9985, 1]$ with mean value
of $0.9997$. For comparison, if we use only one sample ($\omega=1, \theta=1$) for training to obtain a control law, the
testing performance gives fidelities that lie in $[0.9436, 1]$ with mean value
of $0.9808$.

These numerical results show that SLC is effective for control
design of the two-level inhomogeneous ensemble. The fidelities of
the controlled state for the randomly
selected members can approach very near to $1$ even with $\pm
20\% $ of parameter dispersion with a uniform distribution on all
the parameters.

\begin{figure}
\centering
\includegraphics[width=3.5in]{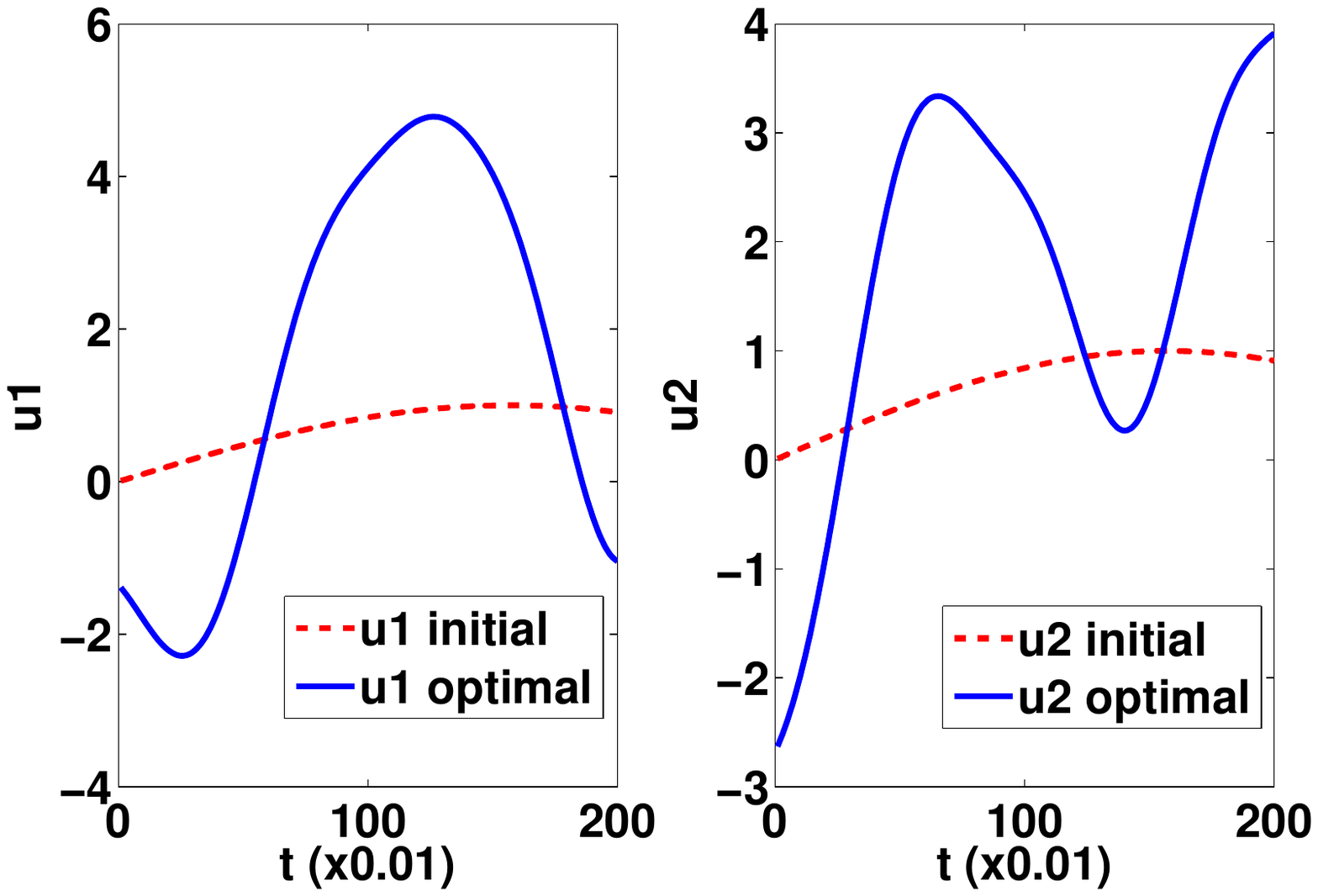}
\caption{(color online) The learned optimal control strategy that maximizes $J_{N}(u)$
for the two-level ensemble with two controls.}
\end{figure}

\begin{figure}
\centering
\includegraphics[width=3.5in]{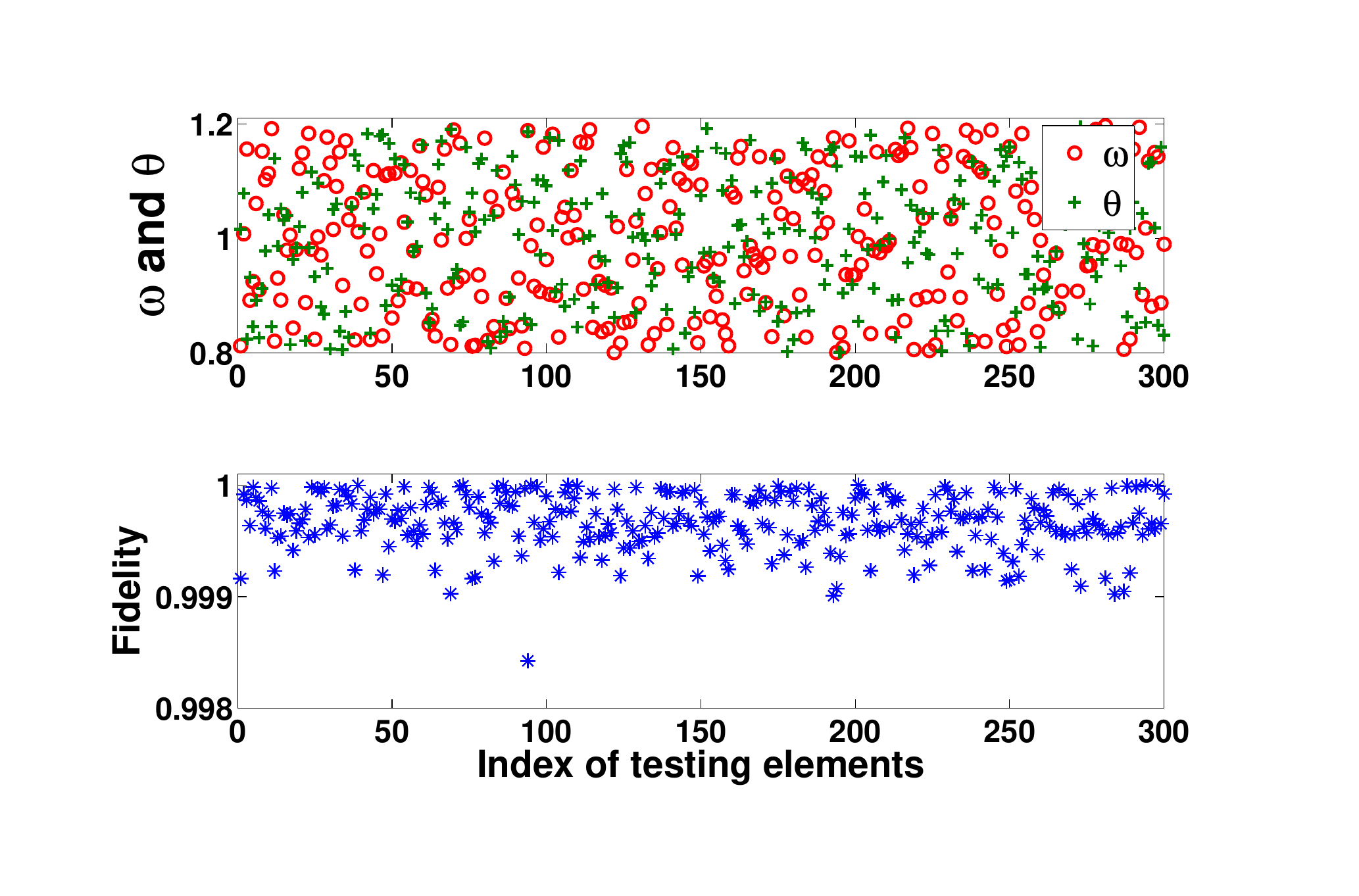}
\caption{(color online) The testing performance (with respect to fidelity) of the
learned optimal control strategy for the two-level ensemble with
two controls.}
\end{figure}

\begin{figure}
\centering
\includegraphics[width=3.5in]{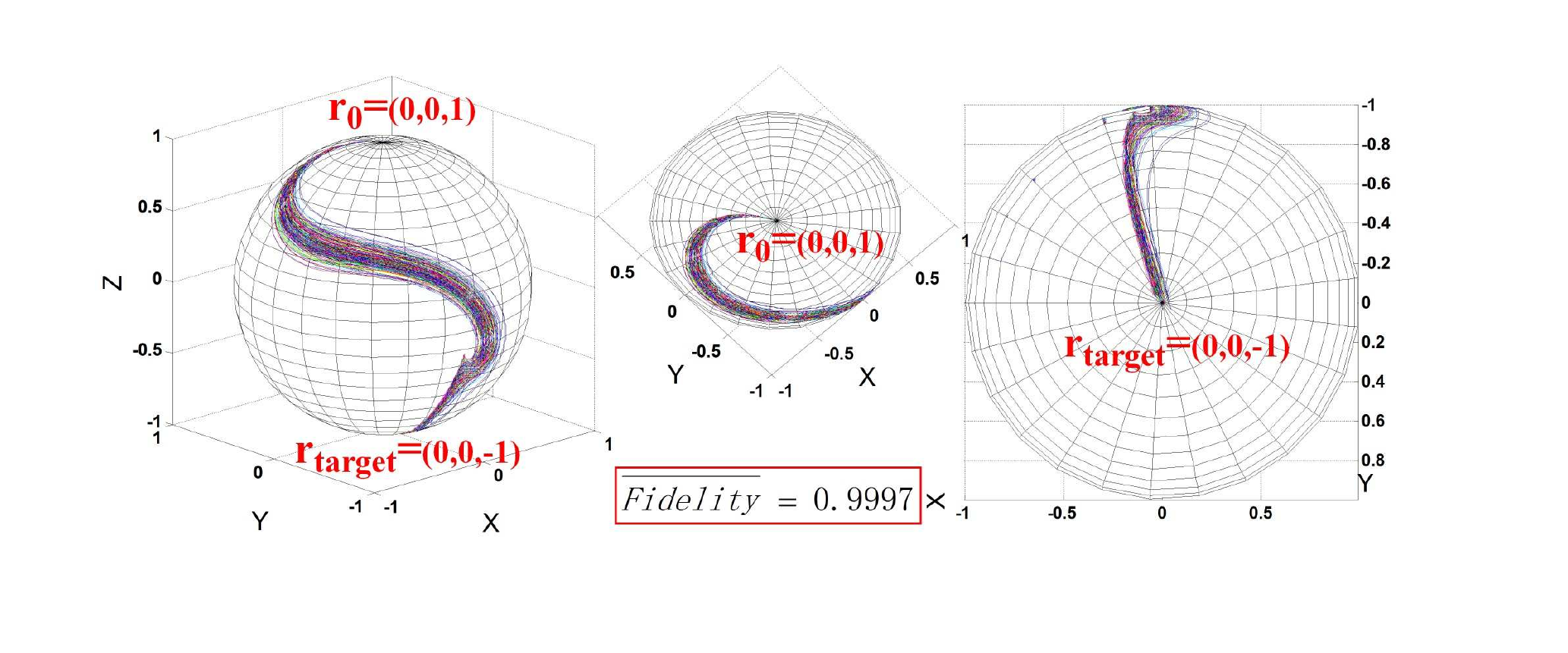}
\caption{(color online) Demonstration of the state transition trajectories of the
two-level ensemble with inhomogeneities on both $\omega$ and
$\theta$ using the learned optimal control strategy as shown in
Fig. 2 for the state transition problem of guiding from the initial state of
$r_{0}=(0,0,1)$ (i.e., $|\psi_{0}\rangle=|0\rangle$) to the target
state of $r_\textrm{target}=(0,0,-1)$ (i.e.,
$|\psi_\textrm{target}\rangle=|1\rangle$).}
\end{figure}

Using the optimal control strategy in Fig. 2, we randomly
select several thousand members and present the state
transition trajectories of the two-level ensemble on the Bloch sphere. For a two-level system on the
Bloch sphere, its state can be represented using a
vector $\mathbf{r}=(x,y,z)$ where $x=\text{tr}\{|\psi\rangle
\langle\psi|\sigma_{x}\}$, $y=\text{tr}\{|\psi\rangle
\langle\psi|\sigma_{y}\}$, $z=\text{tr}\{|\psi\rangle
\langle\psi|\sigma_{z}\}$. As shown in Fig. 4, although the
trajectories of these randomly selected members considerably
differ from each other due to the inhomogeneity of the
ensemble, they are all successfully driven from the initial
state $|\psi_{0}\rangle=|0\rangle$ (i.e., $r_{0}=(0,0,1)$) to the
same target state $|\psi_\textrm{target}\rangle=|1\rangle$ (i.e.,
$r_\textrm{target}=(0,0,-1)$) with high fidelities indicated above.

\begin{figure}
\centering
\includegraphics[width=3.5in]{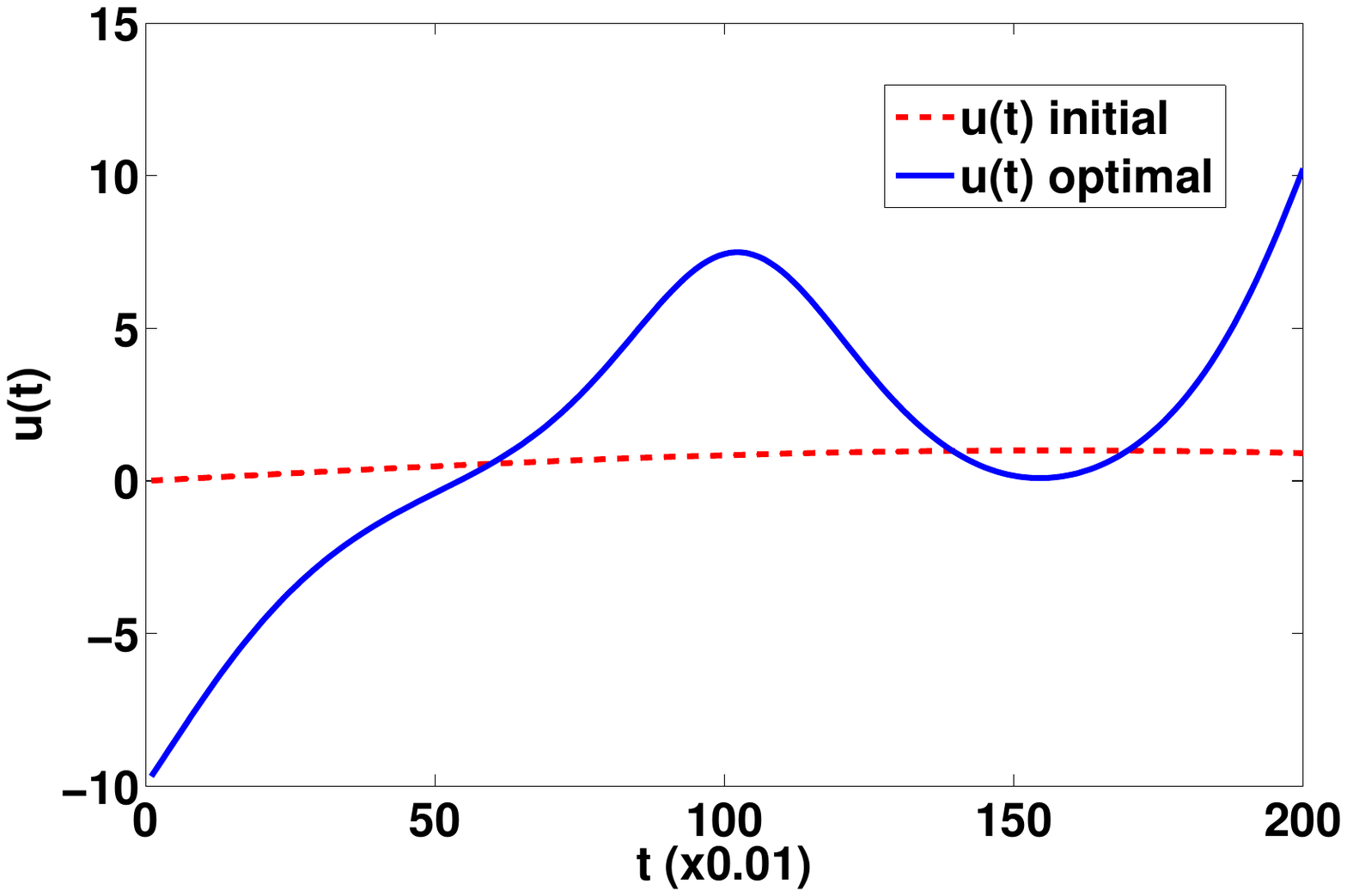}
\caption{(color online) The learned optimal control that maximizes $J_{N}(u)$ for the
two-level ensemble with $u_1(t)=u_2(t)=u(t)$.}
\end{figure}

\begin{figure}
\centering
\includegraphics[width=3.5in]{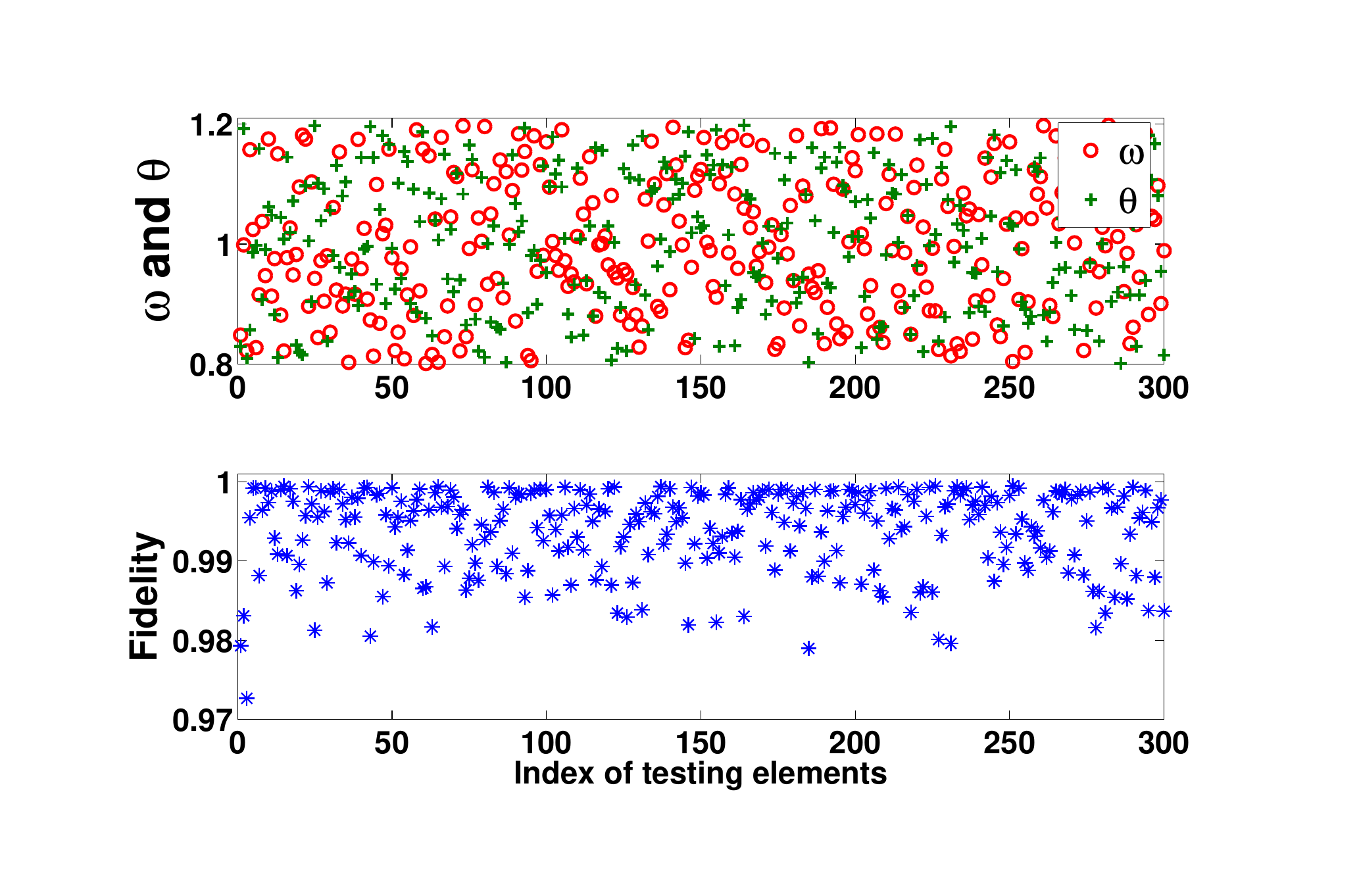}
\caption{(color online) The testing performance (with respect to fidelity) of the
learned optimal control for the two-level ensemble with $u_1(t)=u_2(t)=u(t)$.}
\end{figure}

\subsubsection{Performance with $u_1(t)=u_2(t)$}
Here we consider only one control, i.e., let
$u(t)=u_1(t)=u_2(t)$ with the evolution equation of the ensemble
being
\begin{equation}\label{onecontrol_ensemble2level}
\left(%
\begin{array}{c}
  \dot{c}_{0}(t) \\
  \dot{c}_{1}(t) \\
\end{array}%
\right)=
\left(%
\begin{array}{cc}
  0.5\omega i & \theta h(u) \\
  -\theta h^{*}(u) & -0.5\omega i \\
\end{array}%
\right) \left(%
\begin{array}{c}
  c_{0}(t) \\
  c_{1}(t) \\
\end{array}%
\right),
\end{equation}
where $h(u)=u(t)-0.5iu(t)$. We apply the
same SLC design method and parameter settings as in Section III.B.1 except that $\epsilon=2.0\times 10^{-2}$.
The optimal control
strategy is shown in Fig. 5, and Fig. 6 gives the testing
performance of $300$ randomly selected testing members, whose
fidelities lie in $[0.9727, 1]$ with mean value $0.9939$. Upon comparison with the case in Section III.B.1, the restricted control has reduced the attainable fidelity.

\section{Control of three-level inhomogeneous quantum ensembles}\label{Sec4}
In this section, we further demonstrate SLC with a $\Lambda$-type three-level inhomogeneous
ensemble. We conclude this
section with a summary of the state transition control fidelities
for all the cases in the paper.

\subsection{Control of a $\Lambda$-type atomic ensemble}
We consider a $\Lambda$-type atomic ensemble and
demonstrate the SLC design process. For a $\Lambda$-type atomic
system \cite{Hou et al 2012}, \cite{You and Nori
2011}, we assume that the initial state is
$|\psi(t)\rangle=c_{1}(t)|1\rangle+c_{2}(t)|2\rangle+c_{3}(t)|3\rangle$, and
denote $C(t)=(c_{1}(t),c_{2}(t),c_{3}(t))$ where $c_i(t)$'s are complex amplitudes. We have
\begin{equation}
i\dot{C}(t)=(H_{0}+H_{u}(t))C(t).
\end{equation}
We take $H_{0}=\textrm{diag}(1.5, 1, 0)$ and choose $H_{1}$ and
$H_{2}$ in the control Hamiltonian of Eq. (\ref{controlmodel}) as follows \cite{Hou et al 2012}:
\begin{equation}\label{h0}
H_{1}=
\left(%
\begin{array}{ccc}
  0 & 0 & 0 \\
  0 & 0  & 1 \\
  0 & 1  & 0 \\
\end{array}%
\right), \ H_{2}=
\left(%
\begin{array}{ccc}
  0 & 0 & 1 \\
  0 & 0  & 0 \\
  1 & 0  & 0 \\
\end{array}%
\right).
\end{equation}

To construct a generalized system for the SLC training step, we choose $N$ samples
from the ensemble to form:
\begin{eqnarray}\label{3level-element1}
\left(%
\begin{array}{c}
  \dot{c}_{1,n}(t) \\
  \dot{c}_{2,n}(t) \\
  \dot{c}_{3,n}(t) \\
\end{array}%
\right)=\ \ \ \ \ \ \ \ \ \ \ \ \ \ \ \ \  \ \ \ \ \ \   \ \ \ \ \ \ \ \ \ \ \ \ \ \ \ \ \ \ \ \ \ \ \ \ \nonumber  \\
\left(%
\begin{array}{ccc}
  -1.5\omega_{n} i & 0 & -i\theta_{n} u_{2}(t)  \\
  0 & -\omega_{n} i  & -i\theta_{n} u_{1}(t) \\
  -i\theta_{n} u_{2}(t) & -i\theta_{n} u_{1}(t) & 0 \\
\end{array}%
\right) \left(%
\begin{array}{c}
  c_{1,n}(t) \\
  c_{2,n}(t) \\
  c_{3,n}(t) \\
\end{array}%
\right),
\end{eqnarray}
where $\omega_{n} \in [1-\Omega, 1+\Omega]$ and $\theta_{n} \in
[1-\Theta, 1+\Theta]$ have uniform distributions. The objective
is to find a control strategy $u(t)=\{u_{m}(t), m=1,2\}$ to drive
all the inhomogeneous members from
$|\psi_{0}\rangle=\frac{1}{\sqrt{3}}(|1\rangle+|2\rangle+|3\rangle)$
(i.e.,
$C_{0}=(\frac{1}{\sqrt{3}},\frac{1}{\sqrt{3}},\frac{1}{\sqrt{3}})$)
to $|\psi_{\text{target}}\rangle=|3\rangle$ (i.e.,
$C_{\text{target}}=(0,0,1)$).
We aim to maximize the
performance function
$J_{N}(u)$ in Eq. (\ref{eq:cost}) and employ \emph{Algorithm 1} to find the optimal control
$u^{*}(t)=\{u^{*}_{m}(t), m=1,2\}$ for this generalized system. Then
the optimal control strategy is applied to other randomly selected
members to test its performance.

\subsection{Numerical example}
We use the parameter settings as
follows: the control
strategy is initialized with $u^{k=0}(t)=\{u^{0}_{m}(t)=\sin t,
m=1,2 \}$; $\epsilon=10^{-4}$; the other parameter settings are the same as those of the
numerical experiments for the spin ensemble in Section \ref{Sec3}. To
construct a generalized system for the training step, we have the
training samples selected as follows

\begin{figure}
\centering
\includegraphics[width=3.5in]{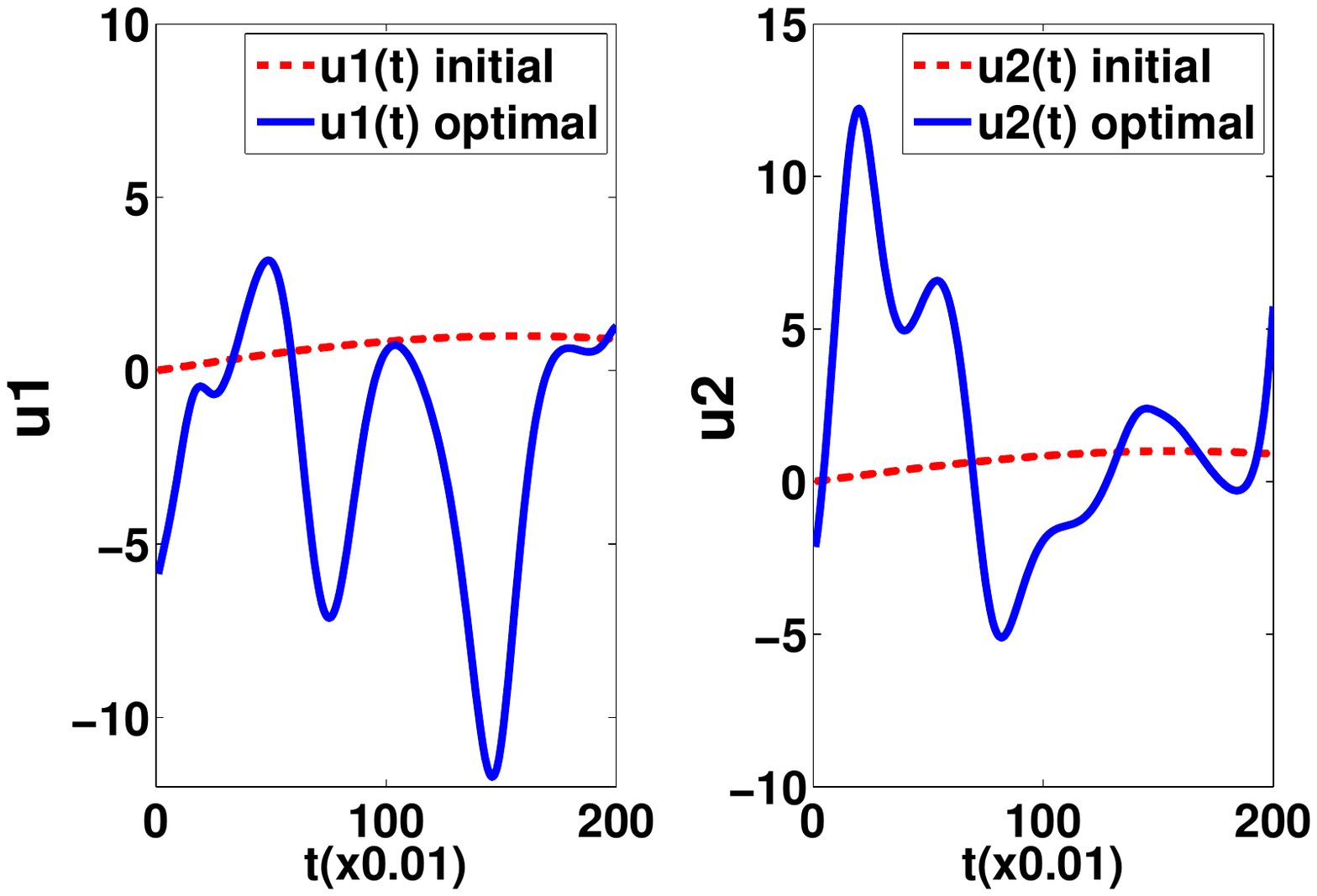}
\caption{(color online) The learned optimal control strategy to maximize $J_{N}(u)$
for the $\Lambda$-type atomic ensemble.}
\end{figure}

\begin{figure}
\centering
\includegraphics[width=3.5in]{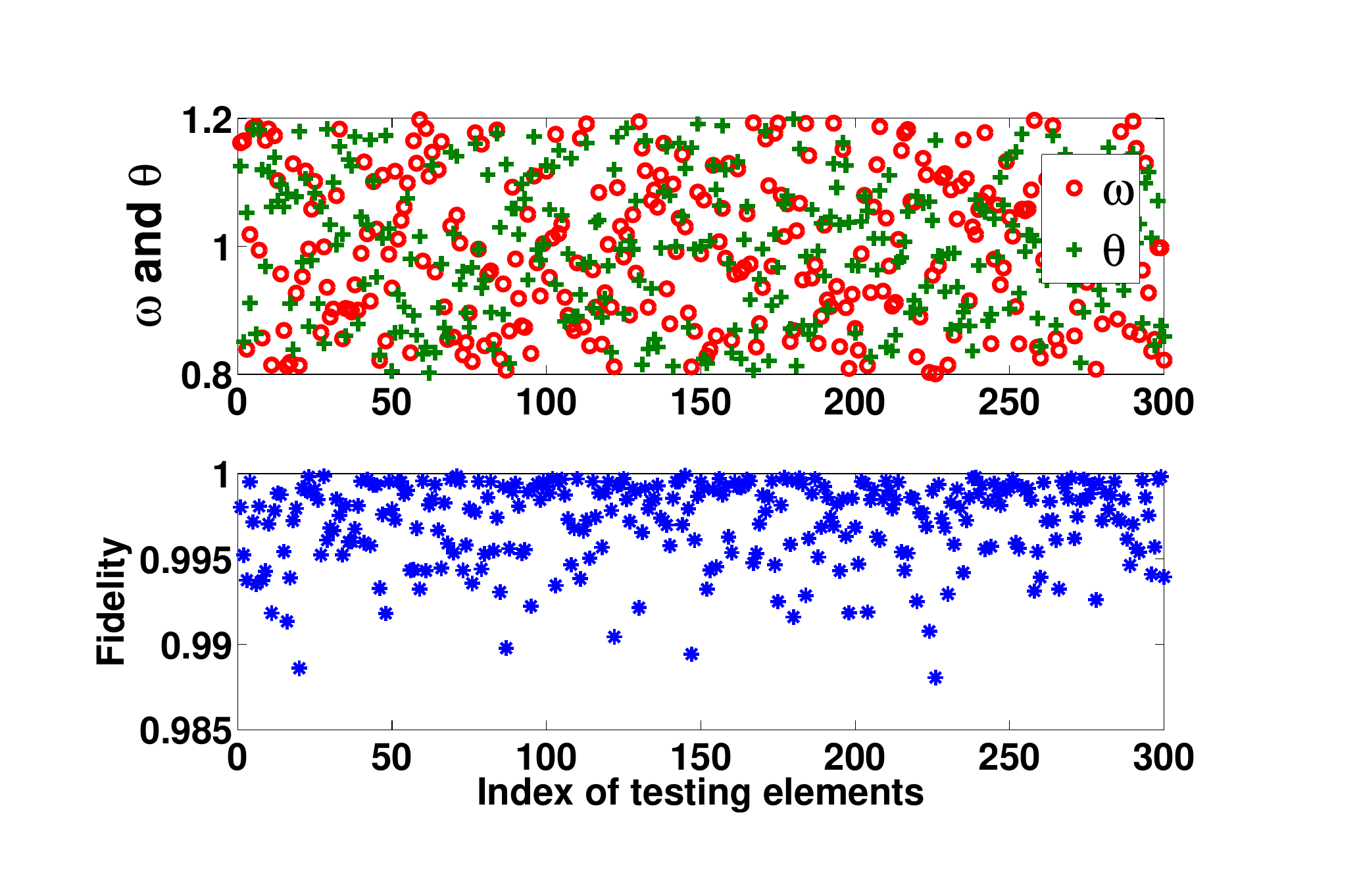}
\caption{(color online) The testing performance (with respect to fidelity) of the
learned optimal control strategy for the $\Lambda$-type atomic
ensemble.}
\end{figure}

\begin{equation}
\left\{ \begin{split}
& \omega_{n}=1-0.2+\frac{0.2(2\text{fix}(n/5)-1)}{5},\\
& \theta_{n}=1-0.2+\frac{0.2(2\text{mod}(n,5)-1)}{5}, \\
\end{split}\right.
\end{equation}
where $n=1,2,\ldots,25$. The learned optimal control strategy is
given in Fig. 7 and the testing results are shown in Fig.
8, which shows the fidelities for all the $300$ testing members lie
in the interval of $[0.9881, 1]$ with the mean value of $0.9972$.
For comparison, if we use only one sample ($\omega=1, \theta=1$) for training to obtain a control law, the
testing performance gives fidelities that lie in $[0.8279, 1]$ with mean value
of $0.9449$.

As a summary of the overall numerical examples, we show the control performance (including some cases that have been explicitly shown above) for the aforementioned spin and $\Lambda$-type atomic
ensembles in Fig. 9. For the two-level inhomogeneous ensemble with
parameter dispersion only in $\omega$, the fidelities of all the 300 testing members
are excellent and lie in the interval of $[1-10^{-6}, 1]$. For the case with parameter dispersion only in $\theta$, the
fidelities lie in the interval of $[0.9987, 1]$ with the mean value 0.9994. If only $u_{1}(t)=u_{2}(t)$ is allowed, the control
performance is not as good as that with two controls $u_{1}(t)$ and $u_{2}(t)$ since we have fewer adjustable degrees of freedom.
All these numerical results further demonstrate previous theoretical predictions that ensemble control should be feasible \cite{Turinici et al 2004JPA}-\cite{Li and Khaneja 2009}.

\begin{figure}
\centering
\includegraphics[width=3.5in]{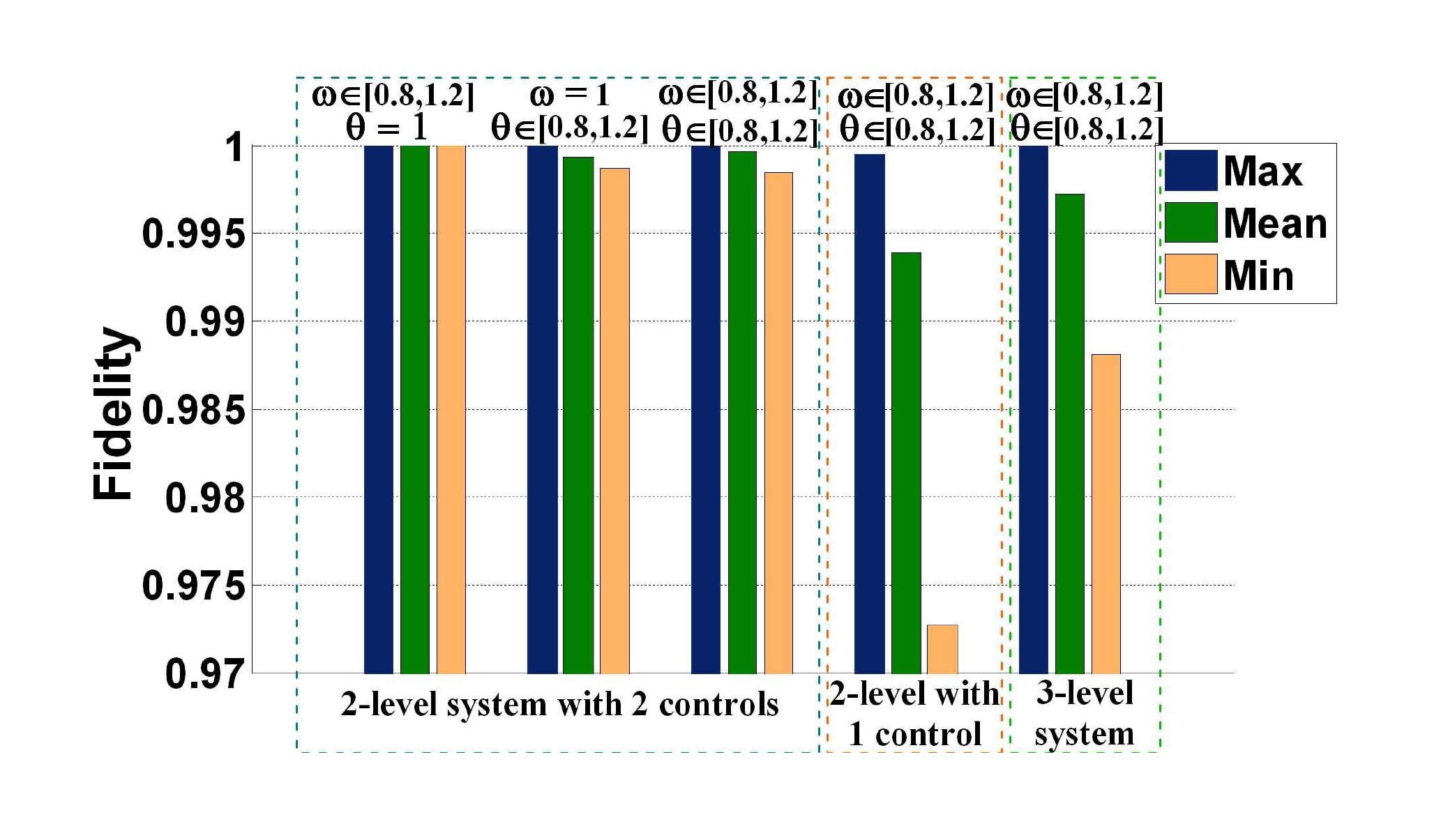}
\caption{(color online) Control performance with respect to fidelity for the
two-level and three-level inhomogeneous ensembles.}
\end{figure}

\section{Conclusions}\label{Sec5}
In this paper, we presented a systematic methodology for control
design of inhomogeneous quantum ensembles. The proposed SLC method
includes the steps of (i) training and (ii) testing.
In the training step, the control is learned for a generalized system
constructed from samples using a gradient flow based
learning and optimization algorithm. In the process of testing, the control
obtained in the first step is evaluated for more randomly selected
members. One approach would be to first learn off-line a control
field using the SLC method, and then generate and apply the control field to an inhomogeneous quantum ensemble in the laboratory.

\section*{Appendix: Gradient flow methods for quantum ensemble control}
To get an optimal control strategy $u^{*}=\{u^{*}_{m}(t), (t \in
[0,T]), m=1,2,\ldots, M\}$ for the generalized system
(\ref{generalized-system}), one technique is to follow the gradient of $J_N(u)$ as an ascent direction. For
ease of notation, we present the method for $M=1$. We introduce a
time-like variable $s$ to characterize different control
strategies $u^{(s)}(t)$. Then a gradient flow in the control space
is defined as
\begin{equation}\label{gradientflowequation}
\frac{du^{(s)}}{ds} =\nabla J_N(u^{(s)}),
\end{equation}
where $\nabla J_N(u)$ denotes the gradient of $J_N$ with respect
to the control $u$. If $u^{(s)}$ is the
solution of \eqref{gradientflowequation} starting from an
arbitrary initial condition $u^{(0)}$, then the value of $J_N$ will
increase along $u^{(s)}$; i.e., $\frac{d}{ds}J_N(u^{(s)})\geq
0$. Starting from a trial guess $u^{0}$, we
solve the following initial value problem
\begin{equation}\label{gradientflowequation2}
\left\{%
\begin{split}
  & \frac{du^{(s)}}{ds} = \nabla J_N(u^{(s)}), \\
  & u^{(0)}=u^{0} \\
\end{split}%
\right.
\end{equation}
in order to find a control strategy which maximizes $J_N$. This
initial value problem can then be solved numerically by using a forward
Euler method (or, if necessary, a high order integration method) over
the $s$-domain; i.e.,
\begin{equation}\label{iteration1}
u(s+\triangle s, t)=u(s,t)+\triangle s\nabla J_N(u^{(s)}).
\end{equation}

In practical applications, in term of a discrete update iteration index $k$,
equation \eqref{iteration1} can be rewritten as
\begin{equation}\label{iteration2}
u^{k+1}(t)=u^{k}(t)+ \eta^{k}\nabla J_N(u^{k}),
\end{equation}
where $\eta^{k}$ is the updating stepsize (learning rate) for the $k\text{th}$ iteration. By Eq. \eqref{eq:cost}, we
also obtain that
\begin{equation}
\nabla J_N(u)=\frac{1}{N}\sum_{n=1}^{N}\nabla J_{\omega_n,\theta_n}(u).
\end{equation} In addition, we have
\begin{equation}\label{gradient-formular}
 \nabla J_{\omega_n,\theta_n}(u)=2\Im\left(\langle\psi_{\omega_n,\theta_n}(T)\vert\psi_{\textrm{target}}\rangle\langle\psi_{\textrm{target}}\vert A_{1}(t)\vert\psi_0\rangle\right)
 \end{equation}
where $A_{1}(t)= U_{\omega_n,\theta_n}(T)U_{\omega_n,\theta_n}^\dagger(t) \theta_n H_1U_{\omega_n,\theta_n}(t)$,
$\Im(\cdot)$ denotes the imaginary part of a complex number, and the propagator $U_{\omega_n,\theta_n}(t)$ satisfies
$$\frac{d}{dt}U_{\omega_n,\theta_n}(t)=-iH_{\omega_n,\theta_n}(t)U_{\omega_n,\theta_n}(t),\quad U(0)=\textrm{Id}.$$
The
gradient flow method can be generalized to the case with $M>1$ as
shown in \emph{Algorithm 1}.

\begin{algorithm}
\caption{Gradient flow based iterative learning}
\label{ModifiedGradientFlow}

\begin{algorithmic}[1]

\State Set the iteration index $k=0$

\State Choose a set of arbitrary controls $u^{k=0}=\{u_{m}^{0}(t),\
m=1,2,\ldots,M\}, t \in [0,T]$

\Repeat {\ (for each iterative process)}

\Repeat {\ (for each training sample member $n=1,2,\ldots,N$)}

\State Compute the propagator $U_{n}^{k}(t)$ with the control
strategy $u^{k}(t)$


\Until {\ $n=N$}

\Repeat {\ (for each control $u_{m}\ (m=1,2,\ldots,M)$ of the
vector $u$)}

\State
$\delta_m^{k}(t)=2\Im\left(\langle\psi_{\omega_n,\theta_n}(T)\vert\psi_{\textrm{target}}\rangle\langle\psi_{\textrm{target}}\vert A_{m}(t)\vert\psi_0\rangle\right)$
where $A_{m}(t)= U_{\omega_n,\theta_n}(T)U_{\omega_n,\theta_n}^\dagger(t) \theta_n H_mU_{\omega_n,\theta_n}(t)$

\State $u_{m}^{k+1}(t)=u_{m}^{k}(t)+\eta^{k} \delta_{m}^{k}(t)$

\Until {\ $m=M$}

\State $k=k+1$

\Until {\ the learning process ends}

\State The optimal control strategy
$u^{*}=\{u_{m}^*\}=\{u_{m}^{k}\}, \ m=1,2,\ldots,M$

\end{algorithmic}
\end{algorithm}

\section*{Acknowledgment}
This work was supported by the Natural Science
Foundation of China (No.61273327),
by the Australian Research Council (DP130101658, FL110100020) and by the US DoE (No. DE-FG02-02ER15344), NSF (No. CHE-0718610) and ARO (No. 0025020).

\end{document}